\documentstyle[prl,aps]{revtex}
\input{psfig}
\draft
\begin{document}

\twocolumn[\hsize\textwidth\columnwidth\hsize\csname@twocolumnfalse\endcsname
\title{Localized and Cellular Patterns in a Vibrated Granular Layer}
\author{Lev S.Tsimring$^1$ and Igor S. Aranson$^{2,3}$}
\address{
$^1$Institute for Nonlinear Science, 
University of California, San Diego, La Jolla,
CA 92093-0402 \\
$^2$ Bar Ilan University, Ramat Gan 52900, Israel\\
$^3$ Argonne National Laboratory,
9700 South  Cass Avenue, Argonne, IL 60439 
}
\date{\today}
\maketitle
\begin{abstract}
We propose a phenomenological model for pattern formation in a
vertically vibrated layer of
granular material. This model exhibits a variety of stable 
cellular patterns including standing rolls and squares as well as 
localized objects ({\it oscillons} and {\it worms}), similar to recent 
experimental observations(Umbanhowar et al., 1996). The model is an
amplitude equation for the parametrical instability coupled to the mass 
conservation law.  The structure and dynamics of the solutions 
resemble closely the properties of localized and cellular 
patterns observed in the experiments. 
\end{abstract}

\narrowtext
\vskip1pc]

Granular materials  exhibit a unique mixture of properties of both 
liquids and solids\cite{jaeger}. 
Intensive theoretical, numerical and experimental studies 
of granular systems revealed a wide variety of new phenomena
typical for granular systems, such as clustering and 
inelastic collapse \cite{collaps}, random force chains
\cite{sue}, granular convection \cite{jaeger1} and  cellular patterns 
in vibrated layers
\cite{swin1,swin2,clement}. The very complicated rheology 
of granular media makes their theoretical
analysis extremely difficult. Unlike fluid dynamics, there is no 
reliable  continuum description of a granular system applicable in a
wide range of conditions. 
The widely used approach is a straightforward 
simulation of many interacting  
particles in a gravity field\cite{ristow,luding}. 

Vibrated granular systems often manifest fluid-like behavior
which resembles similar phenomena in conventional liquids. 
Recent experimental studies of vertically vibrated granular systems 
\cite{swin1,swin2,clement} demonstrated a rich variety of collective
behavior ranging from standing 
waves, hexagons, squares to  localized  objects (particle-like 
{\it oscillons}\cite{swin2} and one-dimensional {\it worms}, see 
\cite{jaeger}).  Cellular  patterns are remarkably similar to Faraday waves
in fluids\cite{faraday} which have recently been a subject of intensive research
(see, e.g.  \cite{param}). The difference, 
however, is that the primary bifurcation to square patterns and oscillons  
is hysteretic: patterns disappear at less magnitude of the 
plate vibrations than they first appear. 
The localized objects, oscillating  at half the frequency of plate
vibrations $\Omega/2$
on the background of a flat surface oscillating at  
$\Omega$, are observed in slightly subcritical parameter
region for cellular patterns. Although some
features of collective behavior of vibrated granular systems were 
reproduced in simulations of a large
ensemble of inelastically interacting particles\cite{luding}, 
the macroscopic understanding of the dynamics 
is lacking. In this Letter we propose  a simple continuum model
exhibiting phenomenology remarkably similar 
to the experimental observations. Due to its simplicity, it is amenable 
to a comprehensive analysis. Although this
model is not derived from the corresponding microscopic equations of
granular systems, it is instructive for interpretation of experimental 
data and may yield testable predictions.

The model consists of an amplitude equation
for the order parameter $\psi$ coupled to a conservation law for an average 
mass of granular material per unit area (or a local averaged  thickness of the
layer): 
\begin{eqnarray}
\partial_t\psi&=&\gamma\psi^*-(1-i\omega)\psi+(1+ib)\nabla^2\psi
-|\psi|^2\psi-\rho\psi \label{eq1}\\
\partial_t\rho&=&  \alpha \nabla\cdot(\rho\nabla|\psi|^2)+\beta\nabla^2\rho 
\label{eq2}
\end{eqnarray}
Eq.(\ref{eq1}) without the last term is a popular model for   
the parametric instability in oscillating liquid layer
(see \cite{vinals,kiyashko}). The order parameter 
$\psi(x,y,t)$ characterizes local complex amplitude 
of the particle oscillations at the frequency $\omega=\Omega/2$. 
Linear terms in this equation can be derived from the dispersion relation
for parametrically driven granular waves, expanded near frequency $\omega$
and corresponding wavenumber $k$ (here $k=\sqrt{\omega/b}$, parameter 
$b$ must chosen to reproduce the correct wavenumber at a given frequency). 
The term $\gamma\psi^*$
provides parametric driving and leads to the excitation of standing waves.
The term $|\psi|^2\psi$ phenomenologically accounts for 
the nonlinear saturation of oscillations provided in  granular materials  
by restitution.  The last term in Eq.(\ref{eq1})  accounts for 
the coupling of the order parameter to the local average density $\rho$.
As  it is observed experimentally \cite{swin1,swin2}, 
the threshold value of 
the vibration amplitude  $\gamma$ for the parametric 
instability depends on the mean layer thickness\cite{swin1} due to 
an increase of internal energy dissipation in thicker layers. 
Although one should generally expect this term to be $f(\rho)\psi$
with $f(\rho)$  saturating at large $\rho$ (when the thickness of the
layer is larger than the scale of typical perturbations),
we hereafter limit ourselves with the simplest form 
$f(\rho)=\rho$ corresponding to relatively thin layers (proportionality 
constant can be omitted after appropriate scaling of $\rho$).

Eq.(\ref{eq2}) describes the conservation of the granular material
where  $\rho$ is a mass of granular material per unit square averaged over
period of vibrations. Two different physical mechanisms contribute to
the in-plane mass flux.
The first term in (\ref{eq2}) reflects the average particle drift due 
to the gradient of magnitude of high-frequency oscillations. On average, 
particles try to ``escape" from regions of large fluctuations, the
effect analogous to formation of 
so called Chladni figures \cite{chladni}.   
The second term 
describes diffusive relaxation of the inhomogeneous mass 
distribution. We expect the effective ``granular temperature" 
and corresponding diffusion constant 
$\beta$ to be proportional to the energy of plate vibrations, 
then for non-vibrating plate the diffusion constant turns to zero and an 
arbitrary pattern ``freezes". 

Let us  briefly discuss the linear stability properties of Eqs.
(\ref{eq1}-\ref{eq2}). The trivial state $\psi=0,\ \rho=\rho_0$ 
corresponding to a flat layer oscillating  
at the driving frequency $\Omega$ becomes unstable at 
$\gamma^2=\gamma_c^2=$ $ (\omega+b(1+\rho_0))^2/(1+b^2)$ for $\omega b 
>1+\rho_0$ with respect to a periodic 
perturbation with the wavenumber $k_c$ given by  $k_c^2=(\omega b-1-\rho_0)
 $ $ /(1+b^2)$. For $\omega b <1+\rho_0$ spatially uniform perturbations 
with $k_c=0$ become unstable first and the vibration threshold  is 
$\gamma_c^2=1+\rho+\omega^2$. Due to the rotational invariance of 
(\ref{eq1}-\ref{eq2}), waves with all directions grow simultaneously.

Above the threshold, the nonlinear terms in Eqs.(\ref{eq1}-\ref{eq2})
saturate the exponential growth of perturbations and provide 
pattern selection. The problem of pattern selection
requires careful analysis of patterns with 
different symmetries. For $\rho=$const  Eqs. (\ref{eq1}-\ref{eq2})
reduce to a single equation for which it is known that
rolls are the only stable cellular pattern above onset
\cite{vinals}. This has been a serious shortcoming of this
model since square patterns are frequently observed in
Faraday experiments\cite{swin1,param}.
Usually pattern selection depends sensitively on the choice of 
nonlinearity in the model, and some tweaking with non-local 
nonlinearities could remedy 
this problem (see e.g. \cite{cross}). It turns out that within our 
combined model with $\rho$ being a dynamical variable it is not needed, 
squares and rolls  emerge naturally in different parameter regions.


Consider stability of simplest cellular patterns (rolls
and squares) in the framework of Eqs. (\ref{eq1}-\ref{eq2}) close to
the threshold of parametric instability. Although the 
analysis is formally valid for arbitrary rhombic pattern, we restrict ourself
by the squares since they are typically preferred by the symmetry. 
Within the framework of weakly-nonlinear analysis (small supercriticality $
\varepsilon$ defined below) cellular patterns are described by the following 
solution to Eq. (\ref{eq1}):
\begin{equation}
\psi= (A\,{\rm sin}(k_c x) + B\,{\rm sin}(k_c y))\,e^{i\,\phi}+w
\label{psi0}
\end{equation}
where $A(t),B(t)$ are the real amplitudes of two (orthogonal) standing waves,
phase $\phi=$const is given by solution of linearized problem,
and $w$ is a  correction to the solution which
we demand to be small at $\varepsilon \to 0$. Near the threshold
the density $\rho$ is enslaved to $|\psi|^2$, and follows the
quasi-stationary solution of Eq. (\ref{eq2})
$\rho =  \rho_0(t) \exp[ - \eta | \psi|^2 ],\ \eta=\alpha/\beta$.
The function $\rho_0(t)$
can be found from the condition of total mass conservation
$S^{-1}\int \rho dx dy = \mu = const$, $S$ is the total area. 
Substituting $\rho$ into
Eq. (\ref{eq1}) and performing standard
orthogonalization procedure to keep $w$ small, we obtain
the following equations for $A(t),B(t)$
(for simplicity we retain nonlinearity only in two first orders)
\begin{equation}
\dot{A}\! =\! A\! \left[ \varepsilon  \!  +\! 
\frac { \mu \eta-\!3}{4}\!  A^{2} \!+\! \frac{2\! \mu\eta\! -\!3}{2}\! B^{2}  -
\frac{\mu \eta^2}{2} \! (B^2  A^{2} \!+\!2 B^{4})\!\right]
\label{A}
\end{equation}
Equation for $B(t)$ is obtained by the permutation $A \iff B $. 
The supercriticality parameter 
is given by $\varepsilon= \gamma_c (\gamma-\gamma_c)/(1+\mu+k_c^2)$.
\cite{note}.
It follows from Eq.(\ref{A}) that 
hysteretic transition to squares ($A=B$) occurs if $\mu\eta>9/5$  
and stripes ($A \ne 0, B=0$ or $A=0, B \ne 0$) exhibit subcritical 
bifurcation at  $\mu\eta>3$. 

In the supercritical case we can drop 
last two terms in Eq.(\ref{A}).
It is easy to verify that for $\varepsilon \to +0$ the
square pattern ($A=B$) is stable for $\mu \eta >1$ and unstable
otherwise. Rolls, in the limit  $\varepsilon \to +0$, are stable
for $\mu\eta<1$ and unstable otherwise.
For larger $\varepsilon>0$ the higher order terms in 
Eq. (\ref{A}) become important, and squares are stable at 
$16\mu\varepsilon \eta^2<3(10\varepsilon\eta-\varepsilon^2\eta^2-9)$ 
and rolls are stable at $4\mu\varepsilon \eta^2>
3(\varepsilon\eta-\varepsilon^2\eta^2 -3)$. 
In the subcritical case $\eta\mu>9/5$ squares are  stable in their entire 
basin of existence given by the condition $48 \mu \varepsilon \eta^2+
(5 \eta \mu - 9 )^ 2=0$. The phase diagram
for  roll and squares in shown in Fig. \ref{fig1}.
It is qualitatively consistent with the
experimental observations of transition from rolls to squares with
decreasing the driving frequency if one assumes that $\eta\mu$ decreases 
with increase of frequency.  One expects that the relative effect of particle 
drift from regions of intense fluctuations characterized by parameter
$\eta$ diminishes with the increase of $\omega$ since characteristic vertical
scale of the layer involved in oscillations becomes smaller at high frequencies.
At large positive $\varepsilon$ there is a bistable region where 
rolls and squares co-exist, also in agreement with experiments\cite{swin1}).
It should be noted however, that the 
phase diagram of Eq.(\ref{A}) exactly corresponds to the original
model (\ref{eq1}-\ref{eq2}) only in the limit of small amplitude,
otherwise it represents only a Galerkin approximation of the exact solution.
Still, our numerical simulations agreed fairly well with these stability 
limits.


Now we consider the localized solutions.
In experiments\cite{swin2} oscillons appear slightly below 
the threshold of the parametric instability for cellular patterns.
We also found stationary localized axisymmetric solutions 
to Eqs. (\ref{eq1}-\ref{eq2}) in weakly subcritical region
(dots in Fig. \ref{fig1} correspond to stable localized solutions found for various 
combinations of parameters). Fig. \ref{fig2}
shows the radial structure of the order parameter $\psi$ and
corresponding distribution of $\rho$ for such solution. 
This solution corresponds to a dip in the average
mass distribution $\rho=\rho_0 \exp(-\eta |\psi|^2)$.
Due to the symmetry $\psi\to -\psi$, oscillons of opposite polarities 
may coexist in this system. 
It has oscillating 
tails at $r \gg 1$, $\psi(r) \propto r^{-1/2} \exp(p r) $ with the 
(complex) exponent $p$ given by  
$p^2 = -k^2_c+\sqrt{(\gamma^2-\gamma_c^2)/(1+b^2)}$. 
It explains that peaks and craters with elevated periphery 
replace each other on consecutive cycles of plate vibrations.

We studied the linear stability of these localized solution with respect to 
axisymmetric (usually the most dangerous) perturbations.
Some of the results are presented in Fig. \ref{fig3}, where 
we show largest eigenvalue of linearized system $\lambda$,
$|\psi(0)|$ and  the mass deficit 
$m=2 \pi \rho_0 (\int_0^\infty r \exp[-\eta | \psi|^2 ]
dr-1)$. 
The stability region is limited both at large and small $\gamma$ in accord
with experiments. At the edges of the stable region,$\gamma_{c1,2}$,
stable solution corresponding to the oscillon annihilates 
with other unstable localized solutions. 

The interaction of two oscillons can be considered in the spirit
of Ref. \cite{rab}. Since the asymptotic behavior of $\psi$ for  the
oscillon is oscillatory, the interaction force 
$F\sim\mbox{Re}\exp(pr)$
is oscillatory, too. It is natural to expect a variety of bound states.
A numerically found bound states of two oscillons with opposite phases is 
shown in Fig.\ref{fig4}a, and a bound state of four oscillons 
(one positive surrounded by three negative) is shown in Fig.\ref{fig4}b. As in
the experiment\cite{swin2} we found that bound states with coordination numbers
higher than three are unstable.  There also exist a stable bound state of
two like-phased oscillons, but the equilibrium distance 
between the oscillons  is substantially larger than for 
oppositely-phased pairs, resulting in much weaker binding. Small
``granular noise'' probably unbinds such weakly coupled pairs.

We studied the nonlinear evolution of 
oscillons beyond their region of stability  in numerical simulations of 
Eq.  (\ref{eq1}-\ref{eq2}).  Once the lower bound $\gamma_{c1}$ is
passed, the oscillon rapidly decays towards a  trivial state $\psi=0$. 
Increasing $\gamma$ above $\gamma_{c2}$ 
initial spreading lead to a range of different scenarios 
depending on other parameters $\eta,\mu$, $
\omega,\alpha$. For small $\eta\mu$ (supercritical transition) oscillons 
produce a sequence of
concentric rolls.  Depending on $\eta\mu$ they either remain rolls or
break to produce a disordered square pattern. At larger $\eta\mu$ 
oscillon produces other 
oscillons on its periphery as seen in experiments\cite{swin2}. 
It turns out, surprisingly, that following oscillons do not appear 
uniformly around the center but rather organize themselves in 
chains, resembling worm-like patterns (Fig.\ref{fig4}c),
cf. photo by P. Umbanhowar published in \cite{jaeger}. 
We propose explanation of this effect by the conservation of total mass.
Oscillons push the granular material on their periphery. 
Since the excessive mass from oscillons in a chain is re-distributed 
by the diffusion, it spreads more rapidly near the tip of the chain. 
Therefore,  the next oscillon will likely appear there, and the tip
advances (compare with  diffusion-limited growth\cite{dla}). 
This process will continue until  the average
density in the surrounding flat regions becomes so high that the creation 
of new oscillons is halted (the threshold for oscillon stability $\gamma_{c2}$ 
increases with the average density). Our simulations show that 
for values of $\gamma$ slightly above $\gamma_{c2}$ even after a long 
time, oscillons do not fill entire area (see Fig. \ref{fig4}d). 

Let us now return briefly to the problem of selection of cellular patterns. 
In the domain of oscillon stability these patterns can be considered 
as a periodic lattice of weakly-coupled oscillons. As it is well-known
from the solid-state physics, the lowest energy configuration of 
like-charged objects is a hexagonal lattice (compare with Abrikosov 
lattice). In contrast, for  alternatively-charged 
particles the optimal configuration
is a square lattice where a positively-charged particle is surrounded by four 
negatively-charged (anti-ferromagnetic lattice). Eqs.
(\ref{eq1}-\ref{eq2}) preserve the symmetry $\psi\to -\psi$ and therefore
square patterns formed by both positive and negative oscillons dominates
(see Fig.\ref{fig4}d). Once this symmetry is broken (as in two frequency 
driving), patterns are formed by like-phased oscillons, and the hexagonal 
symmetry can be expected. The selection of the pattern in supercritical case 
where oscillons are unstable 
is more subtle. The advantage of the square pattern over
rhombi cannot be determined in
the lowest order of our weakly-nonlinear perturbation theory, which 
happens to be insensitive to the angle between two standing waves. 
Our arguments related to optimal packing of
localized structures suggests that the unique selection of square patterns
should occur in higher orders of $\varepsilon$. 


We have shown on the basis of
phenomenological model that the constraint of 
mass conservation plays a crucial role in pattern formation in
vibrated granular materials. It leads to a subcritical bifurcation
towards planar cellular patterns which naturally include
squares in a wide range of parameters. The model also reproduces
stable localized particle-like excitations (oscillons) and 
chains of oscillons or short rolls sometimes called worms. 
The  parameters of the model
can be estimated experimentally or from molecular dynamics. 
We can speculate that our results are also relevant
for fluids, where square patterns are ubiquitous 
and localized objects were observed recently
\cite{feinb}. Parametric waves on a fluid surface induce mean 
surface displacement which must obey a conservation law. 
Thus, we expect that a coupled set of equations for the
order parameter and a mean displacement may serve as a 
paradigm model for this system. Although for fluids the knowledge 
of the Navier-Stokes equations allows for the direct stability analysis 
of cellular patterns\cite{vinals1}, finding a relevant order-parameter 
model would be useful for studies of more complicated pattern 
formation in this rich experimental system. Worm-like structures have also 
been recently observed in electroconvection\cite{dennin}. It is plausible 
to assume that as in a granular system, this effect can also be 
interpreted as 
the growth in a system exhibiting a first-order transition 
to a cellular structure and controlled by a slowly diffusing field.   

We thank M.I.Rabinovich, M.Mungan, H.L.Swinney, P. Umbanhowar and
H.Levine for 
illuminating discussions.  This work was supported by the U.S. Department 
of Energy under contracts DE-FG03-96ER14592 (LT) and W-31-109-ENG-38 (IA).  
The work of IA was also supported by  NSF, Office of Science and Technology 
Center under contract No. DMR91-20000.

\vspace{-0.9in}
\leftline{\psfig{figure=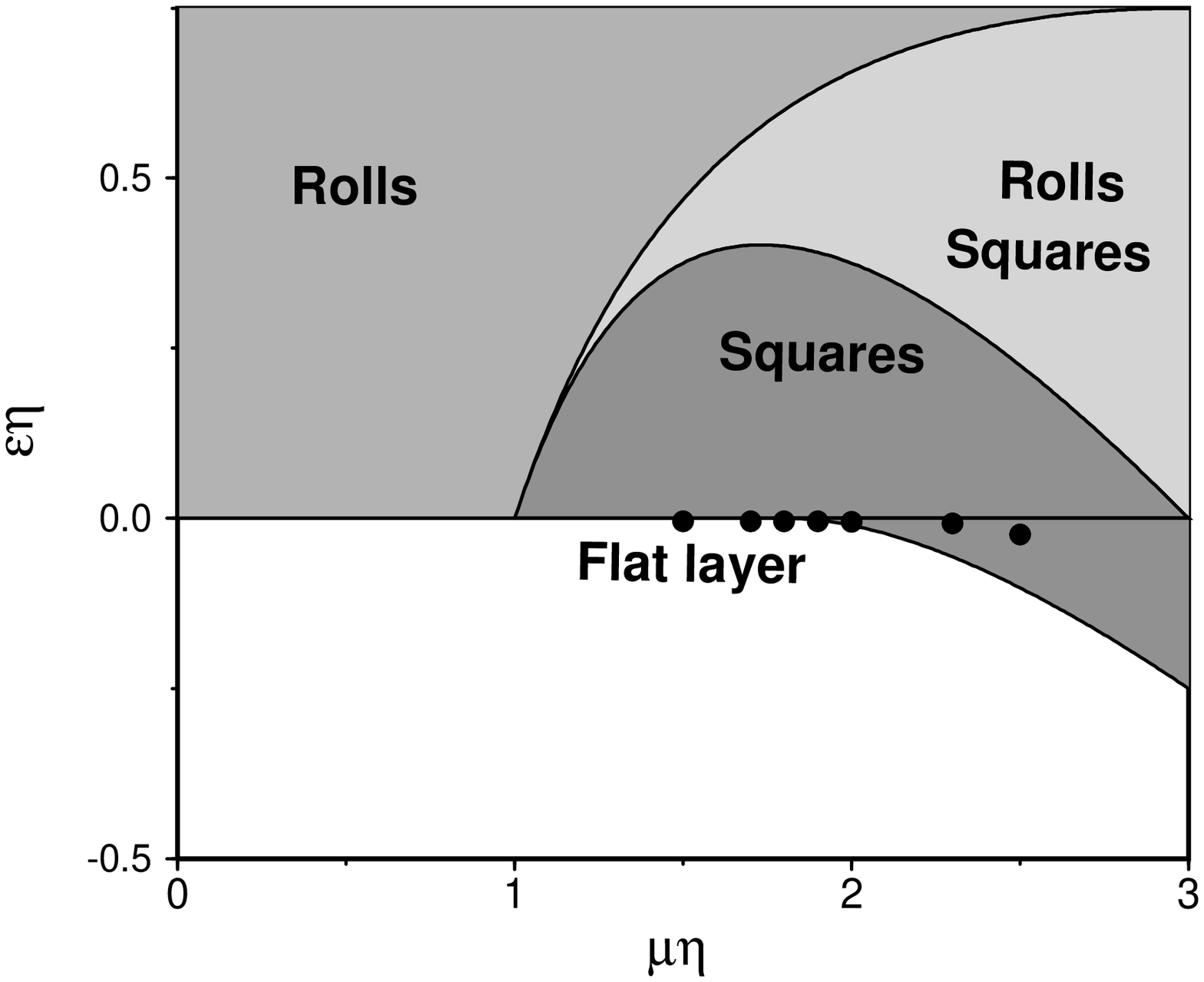,height=2.8in}}
\vspace{-0.2in}
\begin{figure}
\caption{Phase diagram for square and roll patterns, weakly-nonlinear theory. 
Black dots indicate stable oscillons seen in numerical experiments}
\label{fig1}
\end{figure}

\vspace{-0.7in}
\leftline{\psfig{figure=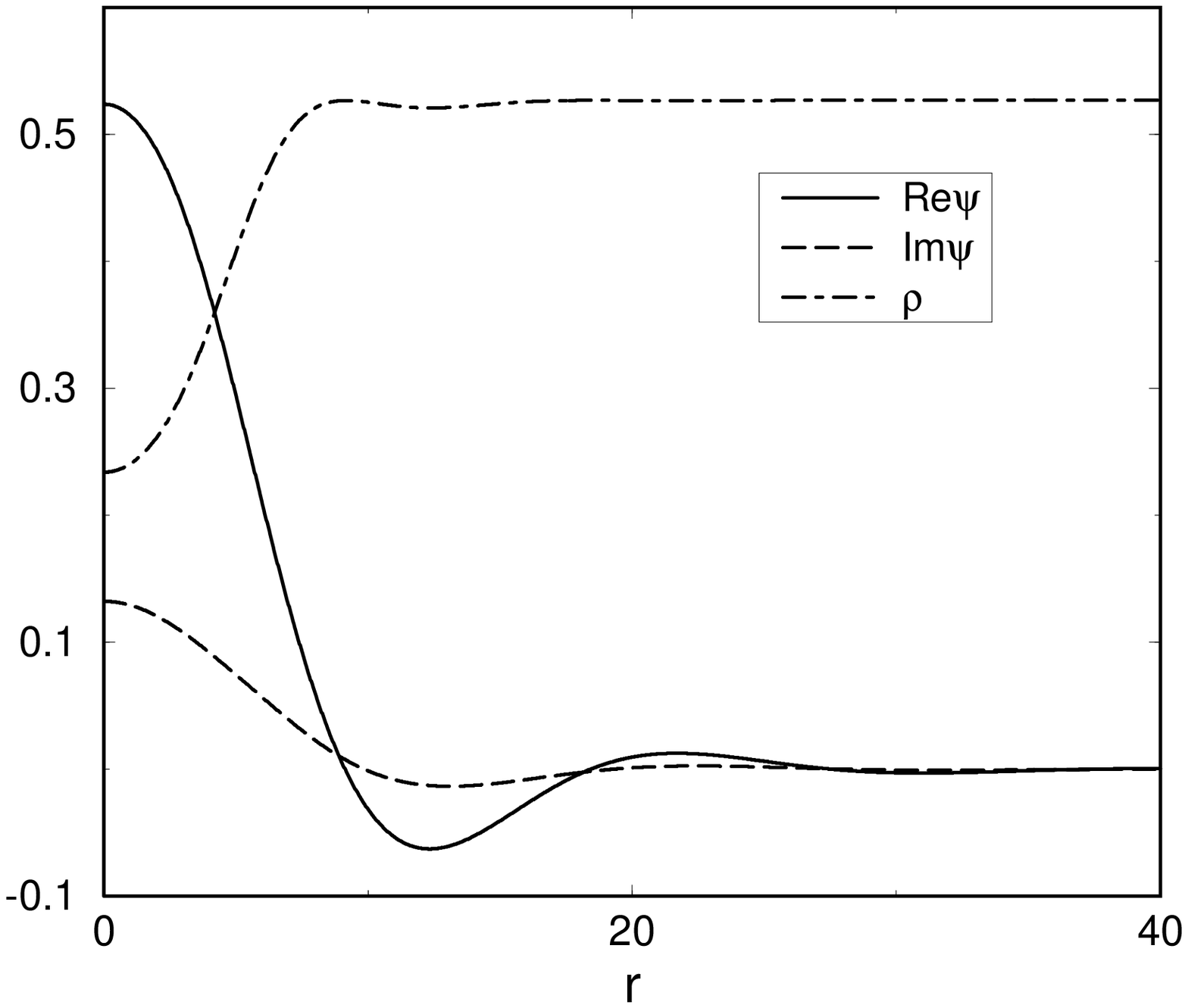,height=2.4in}}
\vspace{-0.3in}
\begin{figure}
\caption{Radial structure of the stable oscillon for 
$\gamma=1.8$, $\mu=0.527$, $b=2,$ $\omega=\alpha=1$, $\eta=2.78$}
\label{fig2}
\end{figure}

\vspace{-1.in}
\leftline{\psfig{figure=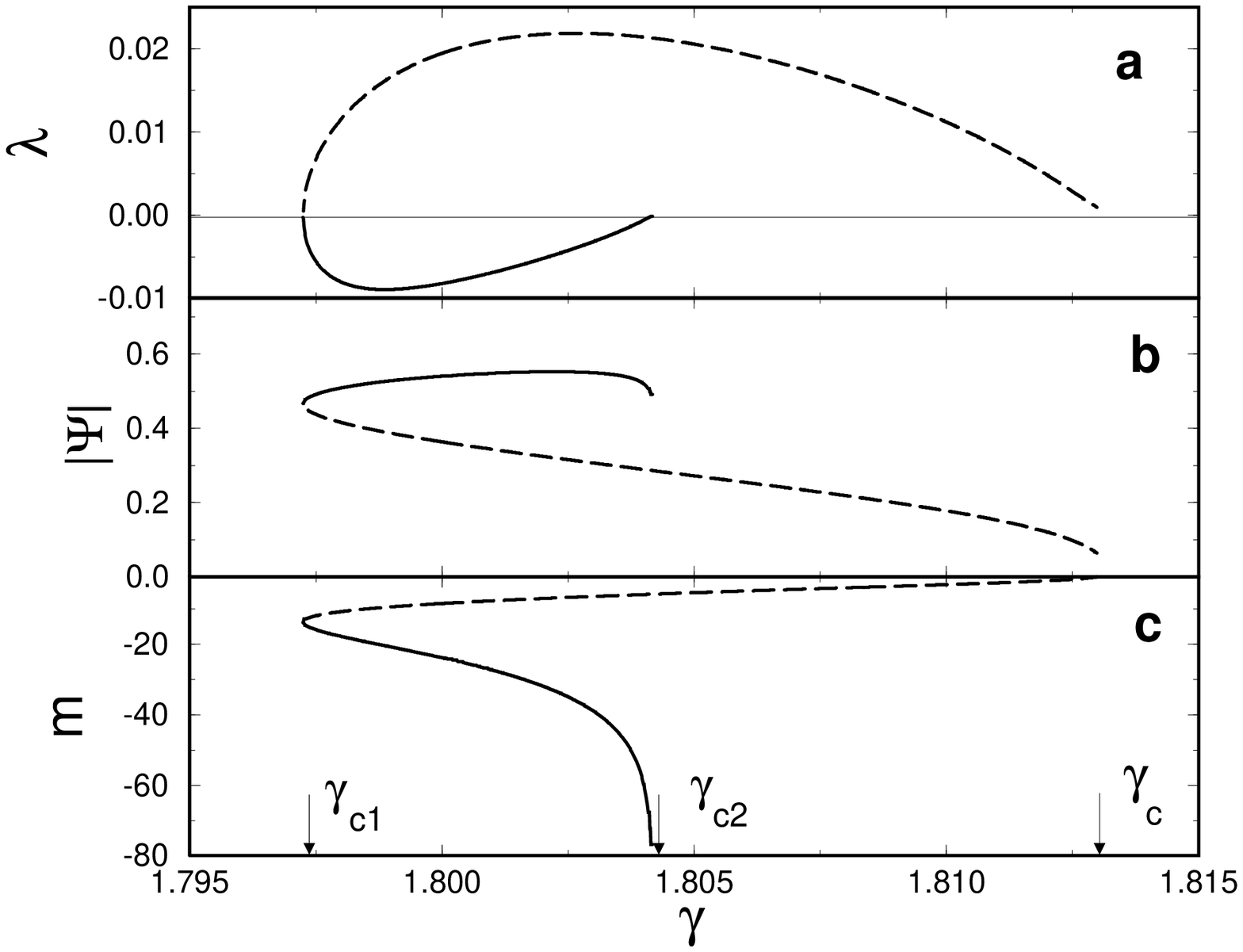,height=2.8in}}
\vspace{-0.3in}
\begin{figure}
\caption{ 
Largest eigenvalue $\lambda$ (a), $|\psi| $ at $r=0$ (b) and 
mass deficit $m$ (c) for stable (solid line) and unstable (dashed line)
oscillons, $\mu=0.527$, $b=2$,$\omega=\alpha=1$, $\eta=$$5/\gamma$.}
\label{fig3}
\end{figure}

\leftline{\psfig{figure=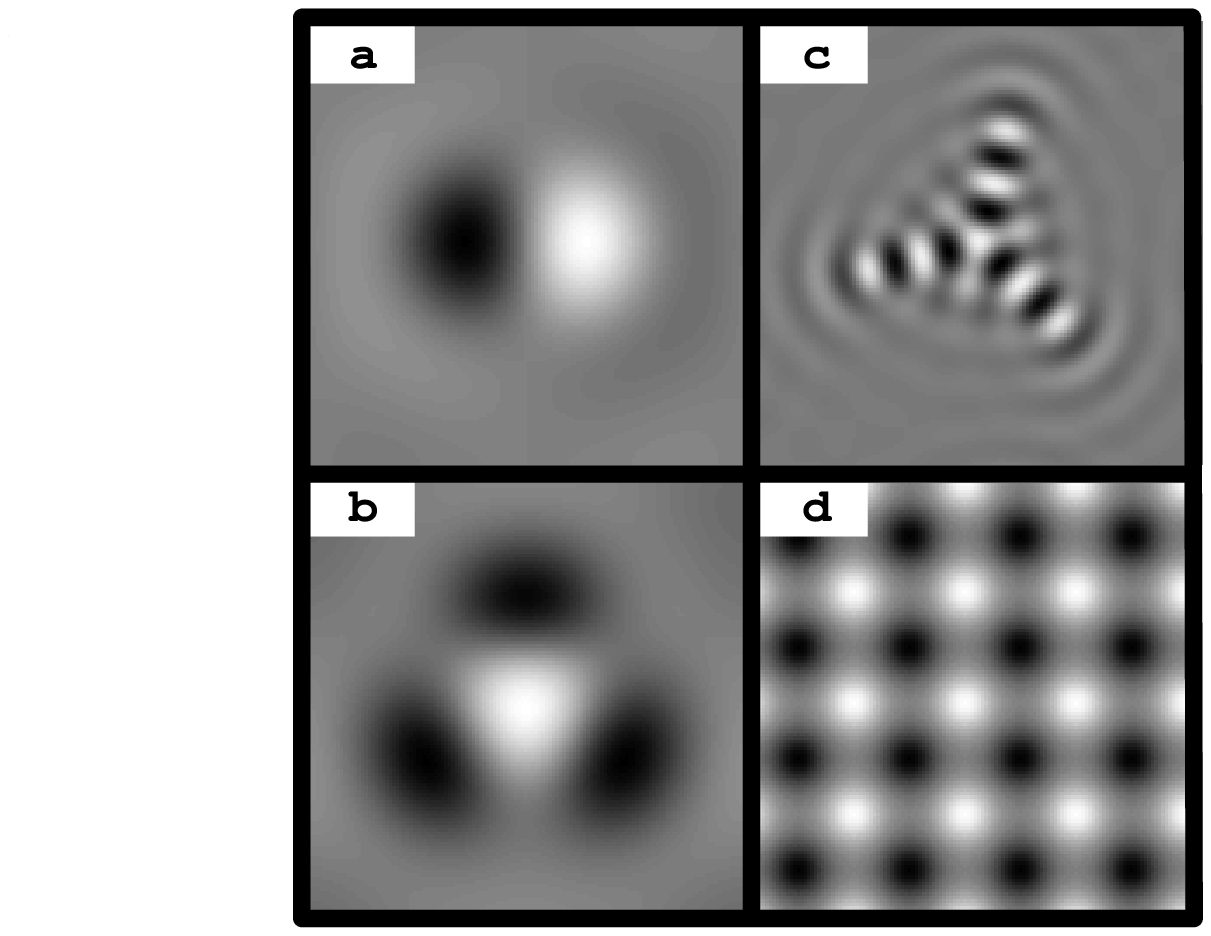,height=2in}}
\begin{figure}
\caption{
Gray-coded images of $\mbox{Re}\psi$ 
(black corresponds to maximum, white to minimum) from
simulations of Eqs. (\protect\ref{eq1}-\protect\ref{eq2}),
(a) bound state of oppositely-phased oscillons,
$\alpha=\omega=1$,$b=2$,$\eta=2.78$,$\mu=0.527$,$\gamma=1.8$,  
size $L=40$;
(b) Triangular bound state, same parameters; 
(c) worm-like structure produced by 
 a single oscillon in the center, $\alpha=1$, $\omega=b=2$,
$\gamma=2.245$,$\eta=4.38$,$\mu=0.525$, $L=100$; 
(d) square lattice, $\omega=\alpha=1$, 
$\gamma=1.84$, $\mu=0.52$, $\eta=2.72$, $L=100$. 
}
\label{fig4}
\end{figure}
\end{document}